# Introduction to eXtended Finite Element (XFEM) Method

## Dibakar Datta

**No. Etudiant : 080579k**

Erasmus MSc in Computational Mechanics

Ecole Centrale de Nantes

FRANCE

Present Address: dibakar_datta@brown.edu or dibdatlab@gmail.com

***Abstract:*** *In the present study the software CrackComput , based on the Xfem and Xcrack libraries has been used for three problems- to experiment on the convergence properties of the method applied to elasto-statics crack problems, comparison of stress intensity factors to simplified analytical results and study of the Brazilian fracture test. All the problems are treated in two dimensions under plane strain assumption and the material is supposed elastic and isotropic. For the first example, comparison for different parameter-enrichment type and radius, degree of polynomial has been performed. Second example convergence of SIF with the L/h ratio has been performed and compared with the analytical solution. Third example is the study of snapback phenomenon.*

## 1. Introduction:

The **extended finite element method (XFEM)**, also known as **generalized finite element method (GFEM)** or **partition of unity method (PUM)** is a numerical technique that extends the classical finite element method (FEM) approach by extending the solution space for solutions to differential equations with discontinuous functions. The extended finite element method was developed to ease difficulties in solving problems with localized features that are not efficiently resolved by mesh refinement. One of the initial applications was the modeling of fractures in a material. In this original implementation, discontinuous basis functions are added to standard polynomial basis functions for nodes that belonged to elements that are intersected by a crack to provide a basis that included crack opening displacements. A key advantage of XFEM is that in such problems the finite element mesh does not need to be updated to track the crack path. Subsequent research has illustrated the more general use of the method for problems involving singularities, material interfaces, regular meshing of micro structural features such as voids, and other problems where a localized feature can be described by an appropriate set of basis functions. It was shown that for some problems, such an embedding of the problem's feature into the approximation space can significantly improve convergence rates and accuracy. Moreover, treating problems with discontinuities with eXtended Finite Element Methods suppresses the need to mesh and remesh the discontinuity surfaces, thus alleviating the computational costs and projection errors associated with conventional finite element methods, at the cost of restricting the discontinuities to mesh edges. The present study is the application of this concept for solving three real life problems.

The outline of the report is as follows. In section 2 the problems of convergence analysis has been described. Section 3 deals with the crack in a beam and comparison of the numerically computed SIF with the analytical one. Section 4 is the study of the Brazilian test. The report is closed in section 5 with some concluding remarks.

# 2. Convergence Analysis:

## 2.1 Problem Statement:

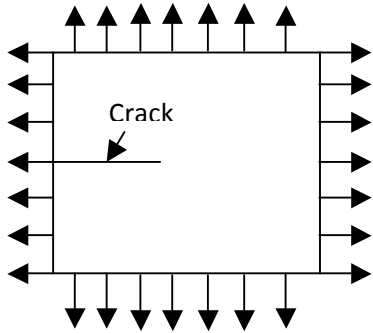

Fig 2.1: Crack in an infinite plane, modeled using stress of the exact solution at the

*Description*: The mode I and II crack opening for an infinite plate will be studied. To emulate the infinite problem, a square shaped domain will be used. On the boundary of the domain, the traction stress of the exact solution is imposed. The elastic numerical displacement field can then be computed numerically on the domain and a H1 norm of the error can be computed in a post processing phase.

*Objective*: The objective of the study is to measure the error between the exact solutions and the numerical solution as well as the convergence rate for different simulation parameters. The improvement related to the use of the tip enrichment function and the size of the enrichment zone is to be studied and the error results are to be presented as curves as a function of element size in log log scale.

## 2.2 Parameters selected for the Problem:

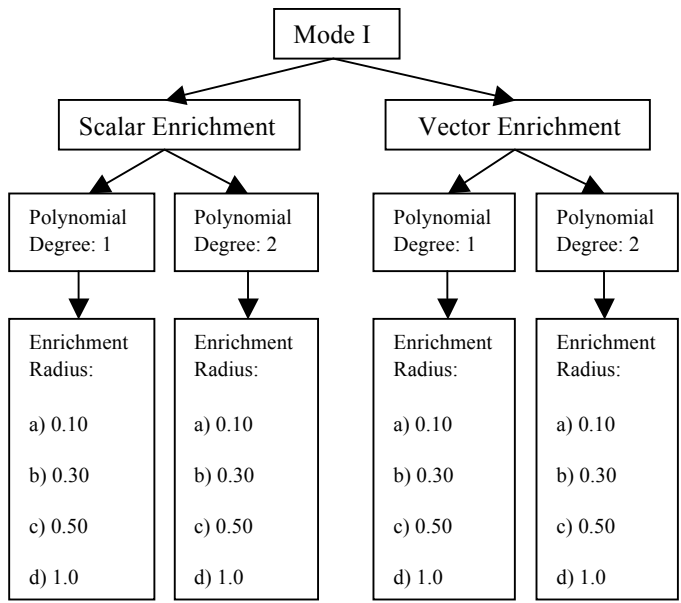

**NOTE:** Simulation performed on a sample size of 10 mm by 1 mm. In each case the simulation is performed using number of elements: 10, 20,30,40,50.

## 2.3 A Brief Theoretical Background:

2.3.1: The concept polynomial in approximation theory: In approximation methods like FEM, the unknown function id approximated as polynomial. When a polynomial is expressed as a sum or difference of terms (e.g., in standard or canonical form), the exponent of the term with the highest exponent is the **degree of the polynomial**. The approximation by of an unknown function by a polynomial will be more close to exact in case a higher order polynomial is used. As shown in the Fig 2.2, the approximation of a quadric polynomial with the piecewise linear function induces error apart from the nodal point. Numerical illustration will show that the selection of higher order polynomial gives less error.

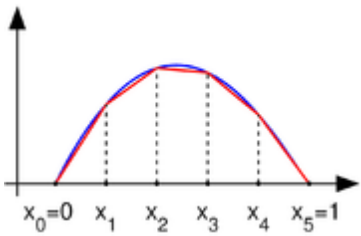 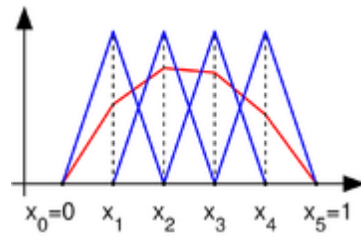 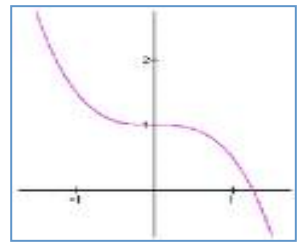 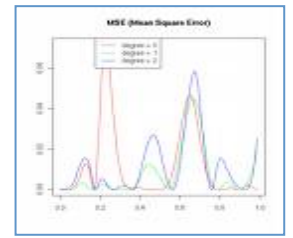

Fig 2.2: A function in $H^1_0$, with zero values at the endpoints (blue), and a piecewise linear approximation (red).

Fig 2.3: Basis functions $v_k$ (blue) and a linear combination of them, which is piecewise linear (red).

Fig 2.4: Second order polynomial. The unknown function is approximated by quadratic polynomial.

Fig 2.5: Higher order polynomial. The unknown function is approximated by cubic, quatric and higher polynomial.

### 2.3.2: The Concept of Enrichment:

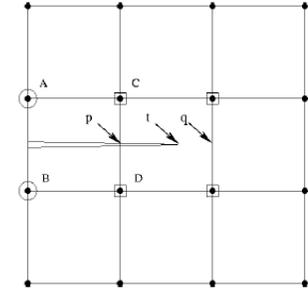

The traditional Finite Element Method (FEM) coupled with meshing tools does not yet manage to simulate efficiently the propagation of 3D cracks for geometries relevant to engineers in industry. In the XFEM approach, In order to represent the crack on its proper length, nodes whose support contains the crack tip (squared nodes shown in figure 2.6) are enriched with discontinuous functions up to the point t but not beyond. Such functions are provided by the asymptotic modes of displacement (elastic if calculation is elastic) at the crack tip.

Fig 2.6: Crack not aligned with a mesh; the circled nodes are enriched with the discontinuous function and the squared nodes with the tip enrichment functions.

The enriched Finite Element approximation is written as:

$$u^h(x) = \sum_{i \in I} u_i N_i(x) + \sum_{i \in I} a_i N_i(x) H(x) + \sum_{i \in K_1} N_i(x) \left( \sum_{l=1}^{4} b_{i,1}^l F_1^l(x) \right) + \sum_{i \in K_2} N_i(x) \left( \sum_{l=1}^{4} b_{i,1}^l F_2^l(x) \right)$$

Where,

- $I$ is the set of nodes in the mesh.
- $N_i$ is the scalar shape function associated to node i.
- $u_i$ is the classical (vectorial) degree of freedom at node i.
- $L \subset I$ is the subset of nodes enriched by the Heaviside function. The corresponding (vectorial) DOF are denoted $a_i$.
- $K_1 \subset I$ and $K_2 \subset I$ are the set of nodes to enrich to model crack tips numbered 1 and 2, respectively. The corresponding degrees of freedom are $b_{i,1}^l$ and $b_{i,2}^l, l = 1,...,4$.
- Functions $F_1^l(x), l = 1,...,4$ modeling the crack tip are given in elasticity by :

$$\{F_1^l(x)\} = \{\sqrt{r}\sin\left(\frac{\theta}{2}\right), \sqrt{r}\cos\left(\frac{\theta}{2}\right), \sqrt{r}\sin\left(\frac{\theta}{2}\right)\sin(\theta), \sqrt{r}\cos\left(\frac{\theta}{2}\right)\sin(\theta)\}$$

*Topological and geometrical enrichment strategies:*

Topological enrichment consists in enriching a set of nodes around a tip. It does not involve the distance from the node to the tip.

Geometrical enrichment consists in enriching all nodes located within a given distance to the crack tip.

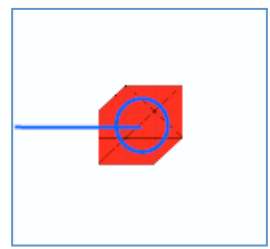 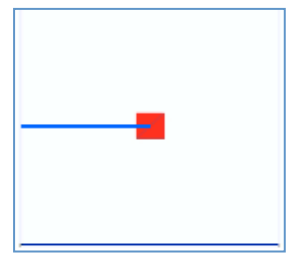

Fig 2.7: Geometrical Enrichment   Fig 2.8: Topological Enrichment

### Vector and Scalar Enrichment (Ref. to Fig 2.6):

Vector Enrichment:

$$\underline{u} = \sum u_i N_i + \sum K_{1i} N_i G_1(r,\theta) + \sum K_{2i} N_i G_2(r,\theta)$$

Scalar Enrichment:

$$\underline{u} = \sum u_i N_i + \sum a_i N_i F_1(r,\theta) + \sum b_i N_i F_2(r,\theta) + \sum c_i N_i F_3(r,\theta) + \sum d_i N_i F_4(r,\theta)$$

## 2.3.3 Analytical Solution:

The numerically computed solution is to be compared with the analytical solution as given below and the H1 norm of the error is to be computed in a post processing phase.

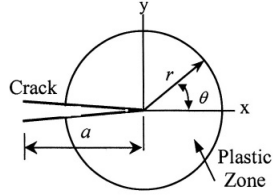

Fig 2.11: Crack tip circular region

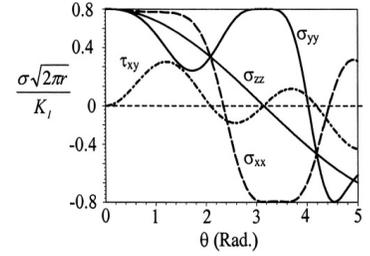

Fig 2.9: Normalized Stress Distribution for Mode 1.

Solution for Stress Field:

$$\begin{bmatrix} \sigma_{xx} \\ \sigma_{yy} \\ \tau_{xy} \end{bmatrix} = \frac{K_I}{\sqrt{2\pi r}} \cos\frac{\theta}{2} \begin{bmatrix} 1 - \sin\frac{\theta}{2}\sin\frac{3\theta}{2} \\ 1 + \sin\frac{\theta}{2}\sin\frac{3\theta}{2} \\ \sin\frac{\theta}{2}\cos\frac{3\theta}{2} \end{bmatrix}$$

Solution for Displacement Field:

$$\begin{bmatrix} \mu_x \\ \mu_y \\ \mu_z \end{bmatrix} = \frac{2K_I}{\sqrt{2\pi E}} \begin{bmatrix} \sqrt{r}\cos\frac{\theta}{2}\left[(1-\upsilon)+(1+\upsilon)\sin^2\frac{\theta}{2}\right] \\ \sqrt{r}\sin\frac{\theta}{2}\left[2-(1+\upsilon)\right]\cos^2\frac{\theta}{2} \\ -\frac{\upsilon B}{\sqrt{r}}\cos\frac{\theta}{2} \end{bmatrix}$$

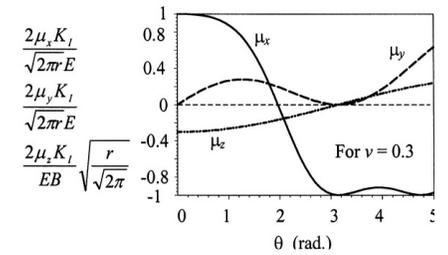

Fig 2.10: Normalized Displacement Distribution for Mode 1.

## 2.3.4: Result and Discussions:

Table 2.1: Table for the error.

| Enrichment type | Radius | Degree | Mode | Error | | | | |
|---|---|---|---|---|---|---|---|---|
| | | | | nelem=10 | nelem=20 | nelem=30 | nelem=40 | nelem=50 |
| Vector | 0.1 | 1 | 1 | 0.244894 | 0.169134 | 0.125591 | 0.106975 | 0.089638 |
| Vector | 0.1 | 2 | 1 | 0.10463 | 0.071143 | 0.0397 | 0.034069 | 0.025796 |
| Vector | 0.3 | 1 | 1 | 0.196668 | 0.130976 | 0.097203 | 0.080513 | 0.068605 |
| Vector | 0.3 | 2 | 1 | 0.068576 | 0.040318 | 0.027794 | 0.022942 | 0.019485 |
| Vector | 0.5 | 1 | 1 | 0.17509 | 0.114009 | 0.085842 | 0.07061 | 0.060568 |
| Vector | 0.5 | 2 | 1 | 0.057539 | 0.033955 | 0.025161 | 0.021125 | 0.018473 |
| Vector | 1 | 1 | 1 | 0.145229 | 0.093322 | 0.071888 | 0.059798 | 0.051911 |
| Vector | 1 | 2 | 1 | 0.047704 | 0.029847 | 0.023623 | 0.020215 | 0.017977 |
| Scalar | 0.1 | 1 | 1 | 0.230946 | 0.151312 | 0.096895 | 0.081758 | 0.063587 |
| Scalar | 0.1 | 2 | 1 | 0.093008 | 0.061212 | 0.029211 | 0.025099 | 0.019488 |
| Scalar | 0.3 | 1 | 1 | 0.143692 | 0.086514 | 0.057819 | 0.045911 | 0.037359 |
| Scalar | 0.3 | 2 | 1 | 0.050444 | 0.030693 | 0.022454 | 0.019104 | 0.016779 |
| Scalar | 0.5 | 1 | 1 | 0.108513 | 0.062699 | 0.043323 | 0.033735 | 0.027722 |
| Scalar | 0.5 | 2 | 1 | 0.043381 | 0.027519 | 0.021656 | 0.018793 | 0.016826 |
| Scalar | 1 | 1 | 1 | 0.056658 | 0.032469 | 0.023866 | 0.019262 | 0.016373 |
| Scalar | 1 | 2 | 1 | 0.038506 | 0.026522 | 0.021816 | 0.019019 | 0.017089 |

### 2.3.4.1: Comparison of error for different enrichment radius:

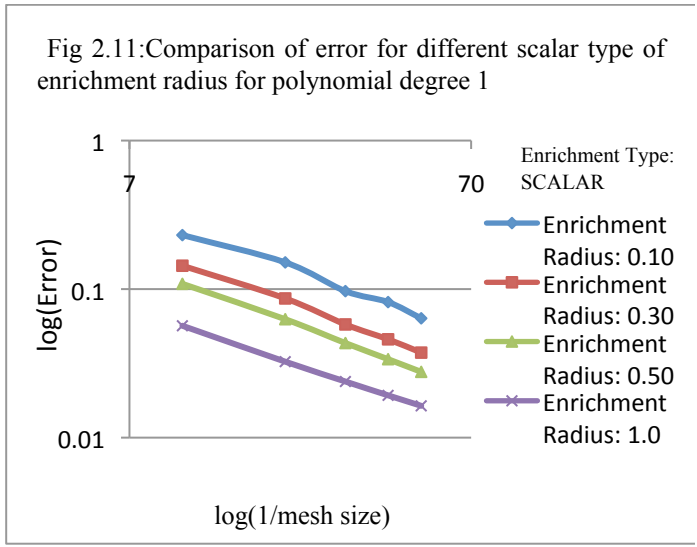

Fig 2.11:Comparison of error for different scalar type of enrichment radius for polynomial degree 1

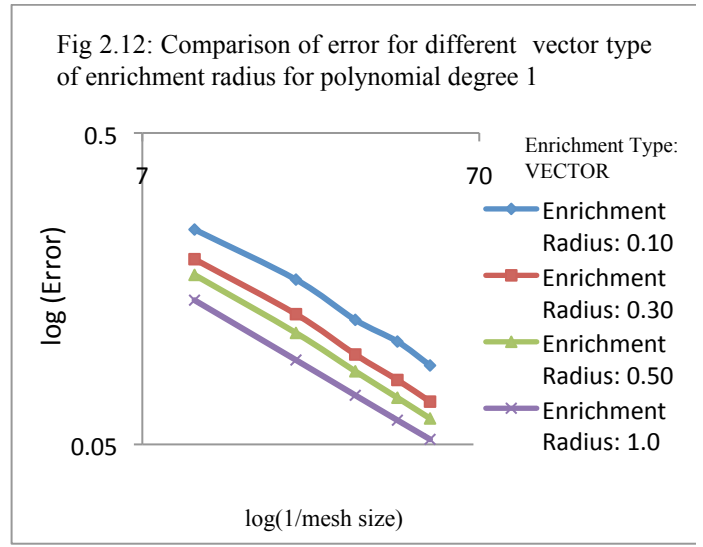

Fig 2.12: Comparison of error for different vector type of enrichment radius for polynomial degree 1

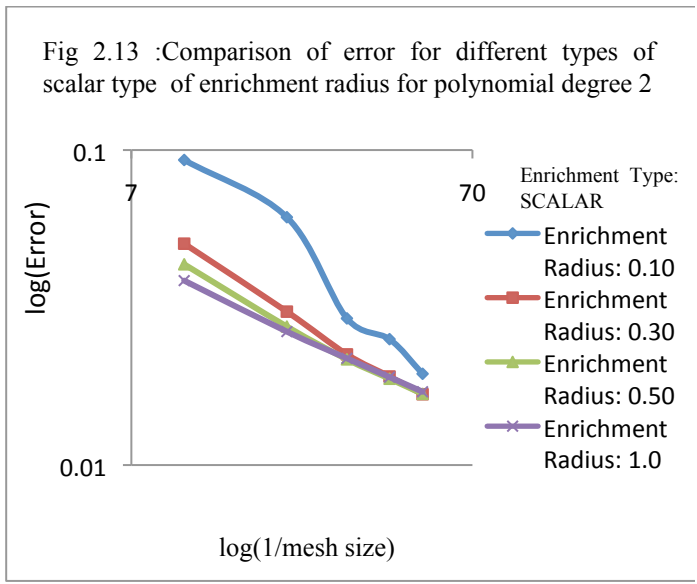

Fig 2.13 :Comparison of error for different types of scalar type of enrichment radius for polynomial degree 2

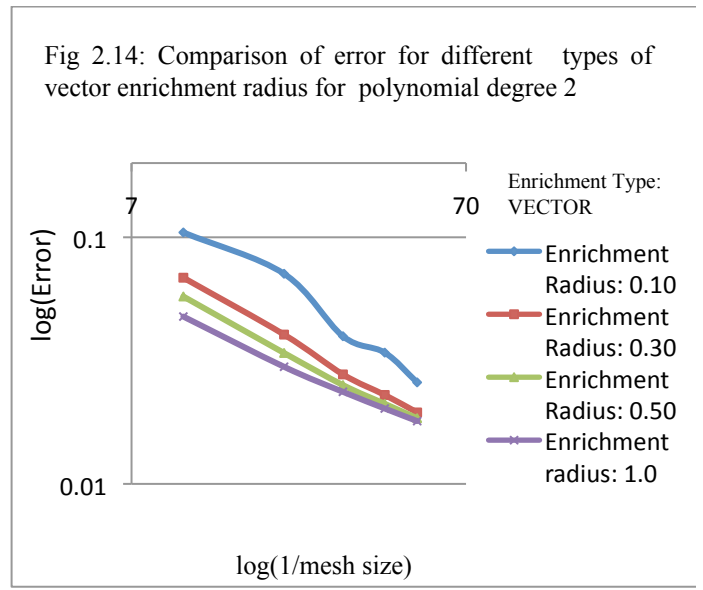

Fig 2.14: Comparison of error for different types of vector enrichment radius for polynomial degree 2

Comment:

- In all four cases, the error due to the enrichment radius 1.0 is less .Because with larger enrichment radius, the number of nodes enriched in the neighborhood of crack tip is more. Hence the approximation function is drawn from the largest space. In general the error can be given by: $\varepsilon = Ch^{\alpha}; h$ is the mesh size. However, in case of traditional FEM approach, with the halving of the mesh size, the error gets reduced by $1/\sqrt{2}$. In case of XFEM, with the conventional topological enrichment, the error gets reduced by ½. Hence with the use of more enrichment function, the reduction of error with the decrease of the mesh size is more.

- The reduction of error with the decrease of mesh size is distinct in case of polynomial degree 1 as in this case the

All the nodes within the specified distance (indicated by blue arrow) from the crack tip are enriched.

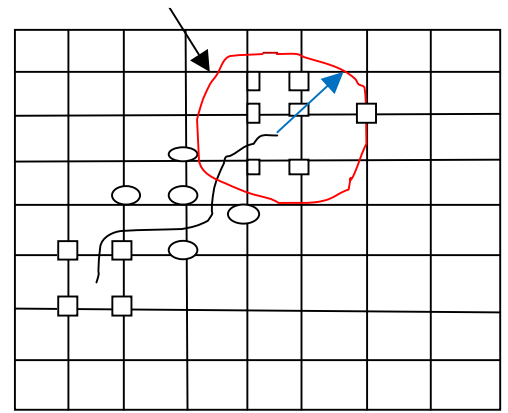

Fig 2.15: Geometric Enrichment

Circled nodes are enriched with the Heaviside function while squared nodes are enriched by tip functions unknown function is approximated with the linear function. Hence *a priori* there is error. Hence the use of more enrichment functions plays a dominant role in reducing the error of approximation.

➢ In case of polynomial degree 2, the reduction of error with the decrease of mesh size is not distinct (especially at the smaller mesh size). Because the use of polynomial degree 2 plays the role of reducing the error. Hence use of higher enrichment radius is of no significant use.

➢ In all cases, the difference of error at larger mesh size is distinct for different enrichment radius. As the error is proportional to the power of h (mesh size). Hence with the smaller mesh size the error due to mesh size is significantly reduced. Hence the reduction of error with the use of higher enrichment radius is not significant.

➢ It is important to note that use of more enrichment function also increases the computation cost. Hence it requires optimizing the enrichment radius in order to avoid the high computation cost.

## 2.3.4.2: Comparison of error for different polynomial degree:

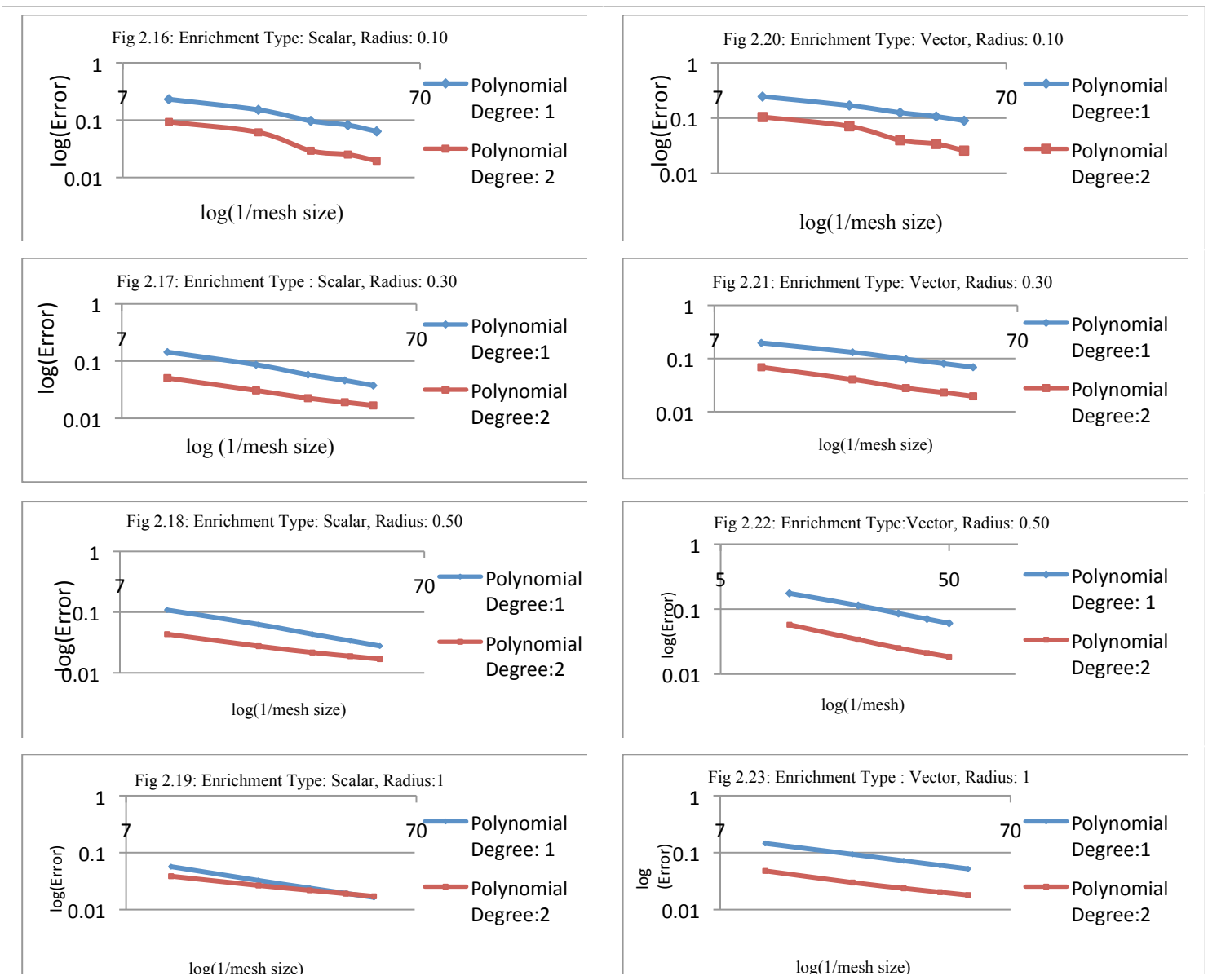

## Comment:

➢ In all the cases, the error is considerably less in case of polynomial degree 2. It is obvious as it can be seen from Fig 2.24 that use of higher order polynomial gives solution close to the exact even with small number of elements as compared to less degree polynomial.

Ref to fig 2.24, quadratic element (polynomial of degree 2) can almost exactly represent an exact solution with just two elements. While the for linear polynomial i.e. polynomial of degree 1, it requires 8 elements.

Hence, for a given number of elements, higher order polynomial gives better result.

Ref. to Fig 2.25, it can be seen that in case of enrichment, higher order makes different.

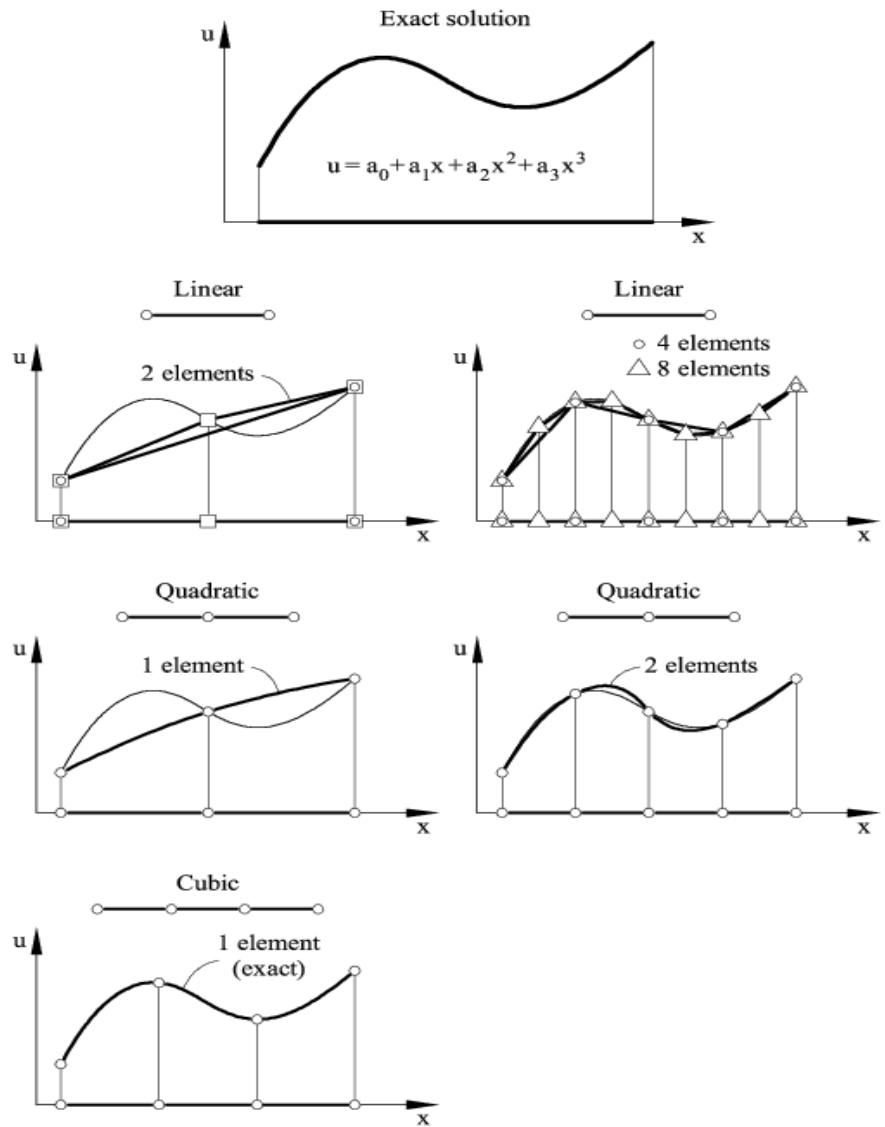

Fig 2.24: Use of different degree polynomial in approximation theory.

➢ It can be observed that for scalar type enrichment with enrichment radius 1, at the smaller mesh size, both polynomial degrees give close result. According to the limited knowledge of the author, reduction of the error mainly governed by scalar type enrichment which uses more number of integration points. This will be thoroughly discussed in the next section.

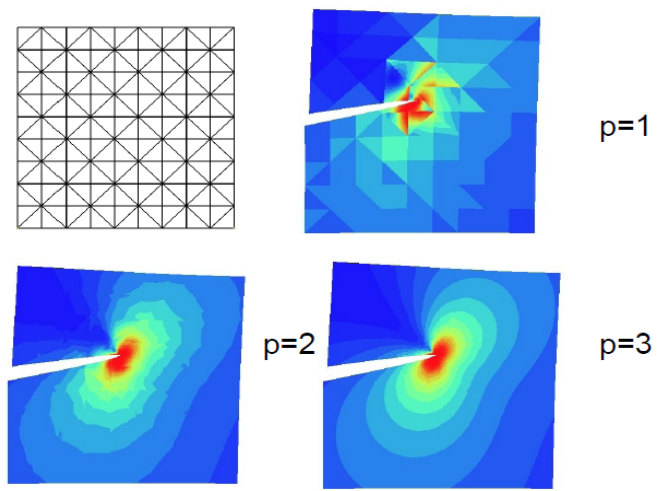

Fig 2.25: In case of enrichment, higher order makes different.

## 2. 3.4.3: Comparison of error for differ

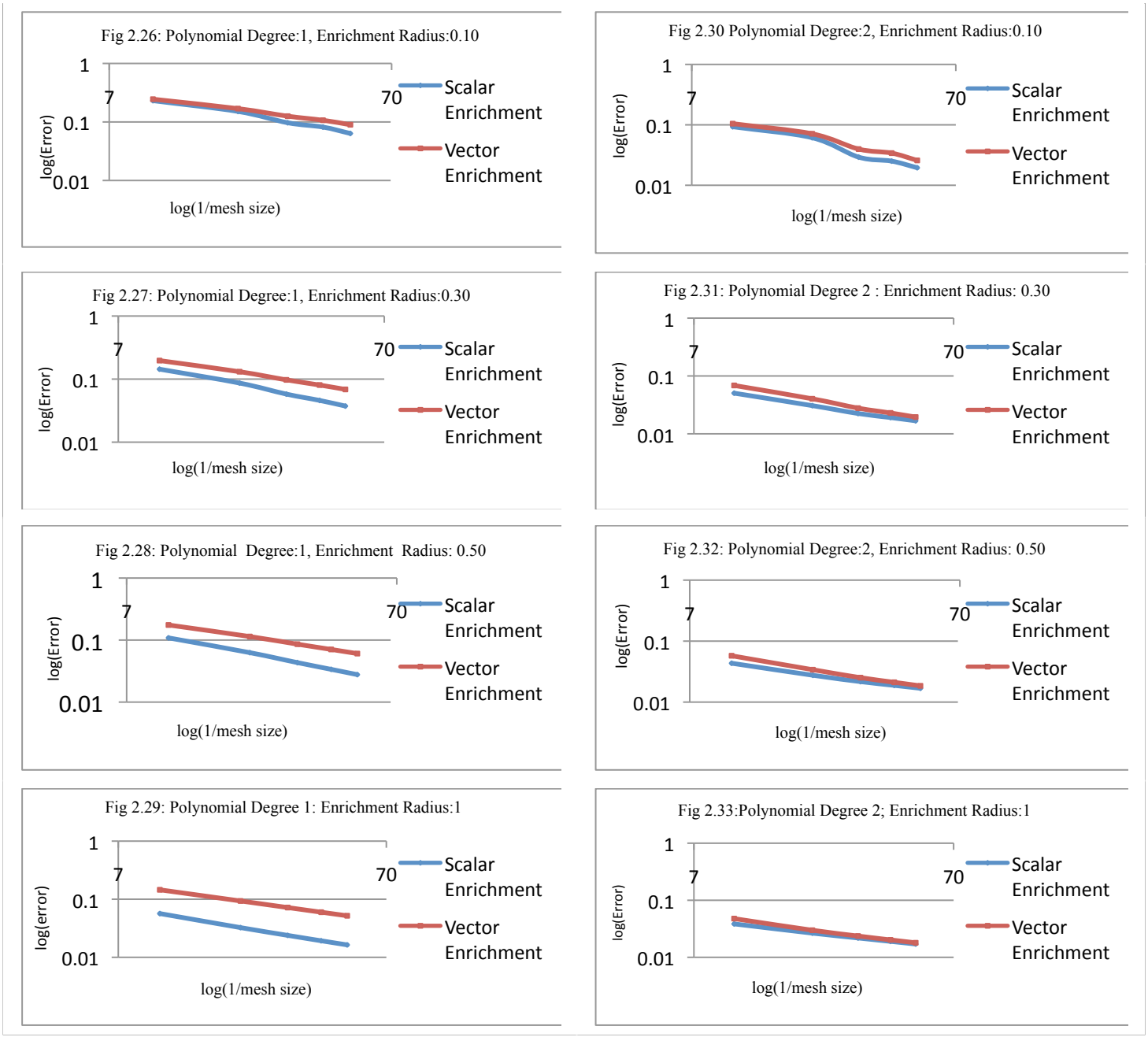

Fig 2.26: Polynomial Degree:1, Enrichment Radius:0.10
Fig 2.30 Polynomial Degree:2, Enrichment Radius:0.10
Fig 2.27: Polynomial Degree:1, Enrichment Radius:0.30
Fig 2.31: Polynomial Degree 2 : Enrichment Radius: 0.30
Fig 2.28: Polynomial Degree:1, Enrichment Radius: 0.50
Fig 2.32: Polynomial Degree:2, Enrichment Radius: 0.50
Fig 2.29: Polynomial Degree 1: Enrichment Radius:1
Fig 2.33:Polynomial Degree 2; Enrichment Radius:1

Comment:

➢ For polynomial degree 1, the error in case of scalar enrichment is considerably less. In scalar enrichment, as mentioned earlier, four enrichment functions are used at each node in two directions. Hence total at each DOF, total 8 DOF are used. Hence more number of integration points is used in this case. In vector enrichment, only 2 DOF (asymptotic mode that needed) is retained and other terms are neglected depending on the 6 coefficients. By playing around with the 4 functions, it exactly represents the function.

$$\underline{u} = \underbrace{K_1 u_1^{aux} N_i + K_2 u_2^{aux} N_i}_{\text{The asymptotic mode that needed}} + \underbrace{\text{Other Terms}}_{\text{Depending on 6 coefficients}}$$

Hence in case of vector enrichment, less number of integration points is used. Hence one of the possible reasons may that use of more number of gauss points for the numerical integration yields better result.

- ➢ In case of polynomial degree 2, error due to scalar and vector enrichment does not differ significantly with the decrease in mesh size. As discussed earlier, higher order polynomial can approximate a function more accurately as compared to the lower order polynomial. Hence for higher order polynomial, the error is not significantly governed by the enrichment type.

2.3.4.4: Displacement and Stress Field:

Displacement Field:

Displacement along the y –direction is given by:

$$\mu_y = \frac{2K_I}{\sqrt{2\pi}E}\left\{\sqrt{r}\sin\frac{\theta}{2}\left[2-(1+v)\right]\cos^2\frac{\theta}{2}\right\}$$

Term causing discontinuity

The displacement field is discontinuous along the crack length.

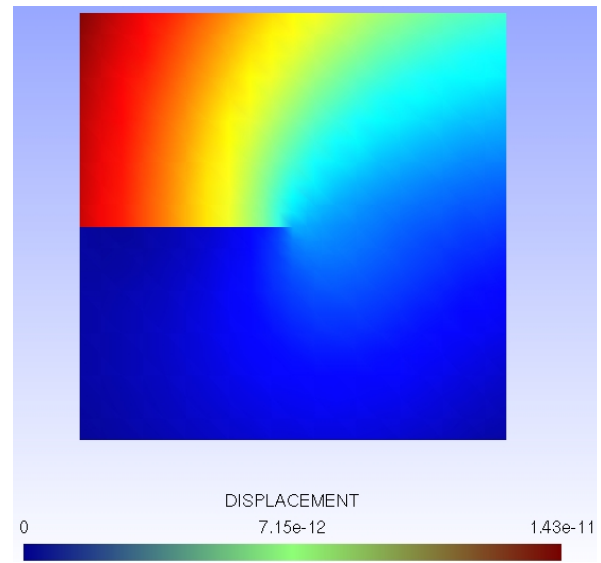

Fig 2.34: Displacement Field.

Stress Field:

As mentioned earlier, the stress field is proportional to $\frac{1}{\sqrt{r}}$. Hence the stress field is singular at the tip of the crack.

At the crack tip, theoretically the stress reaches the maximum value of infinity.

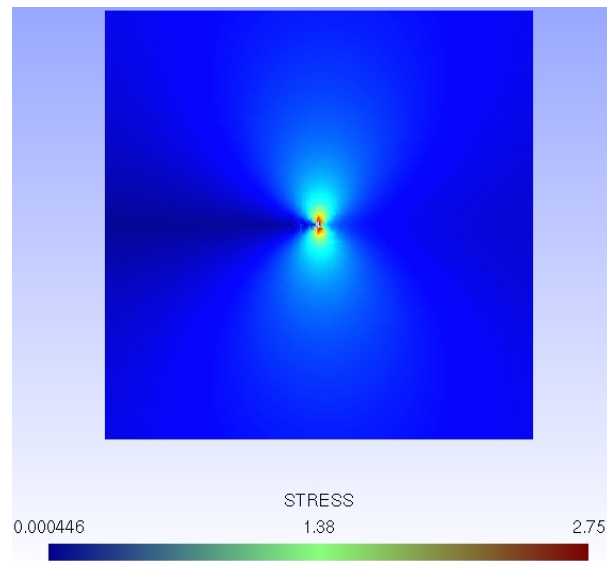

Fig 2.35: Stress Field

# 3. Crack in a beam:

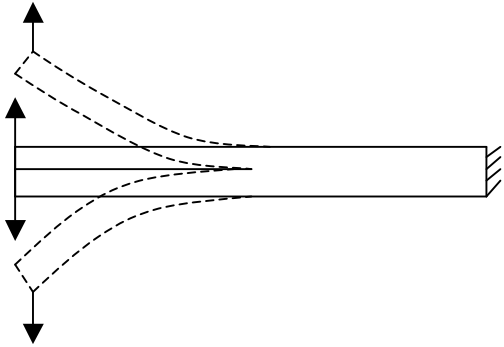

Fig 3.1: Crack in a beam

## 3.1 Problem Statement:

*Description:* A crack in an enhanced beam must be modeled in two dimensions. The stress intensity factor is to be computed for different L/h ratio until convergence.

*Objective:* Comparison and analysis of the analytical stress intensity factor (SIF) with the computed SIF at the crack tip. The analytical model is based on a strain energy analysis on two beams.

## 3.2 Selection of the Mesh Size:

For a particular length, simulation is performed on different mesh size until the stress intensity factor (SIF) for the second mode ($K_{II}$) converges to zero. The following parameter is selected for the analysis.

| Height (h): | 1 |
|---|---|
| Length (L): | 10 |
| Polynomial Degree: | 2 |
| Point on the lip : | 5 |
| Enrichment Radius : | 0.4 |
| Enrichment Type: | Scalar Enrichment |
| Young modulus : | 1 |
| poisson : | 0 |

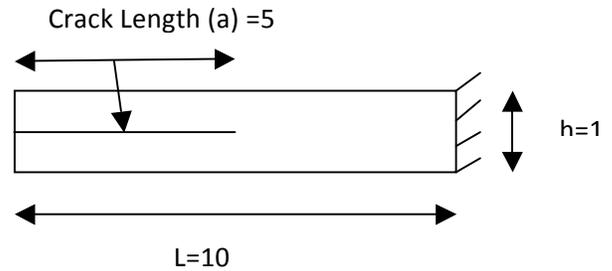

Fig 3.1: Initial geometry for selection of the mesh size.

**NOTE:** The number of element in the longer direction (say M) and in the vertical direction (say N) are selected in such a way so that L/M = h/N.

Table: L v/s $K_{II}$

| L | $K_{II}$ |
|---|---|
| 10 | 1.48E-06 |
| 20 | 3.41E-07 |
| 30 | 2.09E-07 |
| 40 | 1.65E-07 |

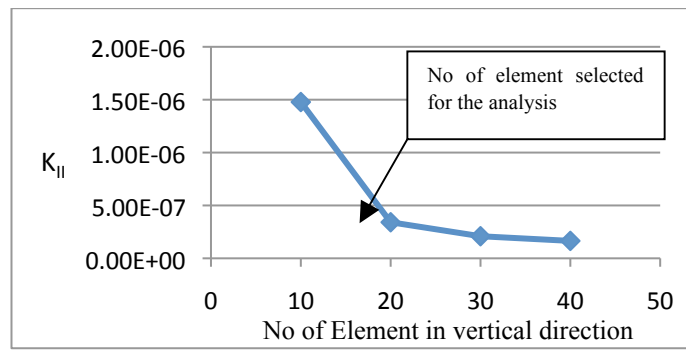

Fig 3.2: No. of element v/s $K_{II}$ plot.

## 3.3 Determination of $K_I$:

The length of the specimen is increased. The length of the crack is kept as half the length of the specimen. The number of element is increased in such a way so that the mesh size in the longer direction is kept constant for all the length.

| L/h | $K_I$ |
|---|---|
| 10 | 5.37E-03 |
| 20 | 1.43E-03 |
| 30 | 6.51E-04 |
| 40 | 3.70E-04 |
| 50 | 2.39E-04 |
| 60 | 1.66E-04 |
| 70 | 1.23E-04 |
| 80 | 9.41E-05 |
| 90 | 7.45E-05 |
| 100 | 6.05E-05 |
| 110 | 5.00E-05 |
| 120 | 4.21E-05 |
| 130 | 3.59E-05 |
| 140 | 3.10E-05 |
| 150 | 2.70E-05 |

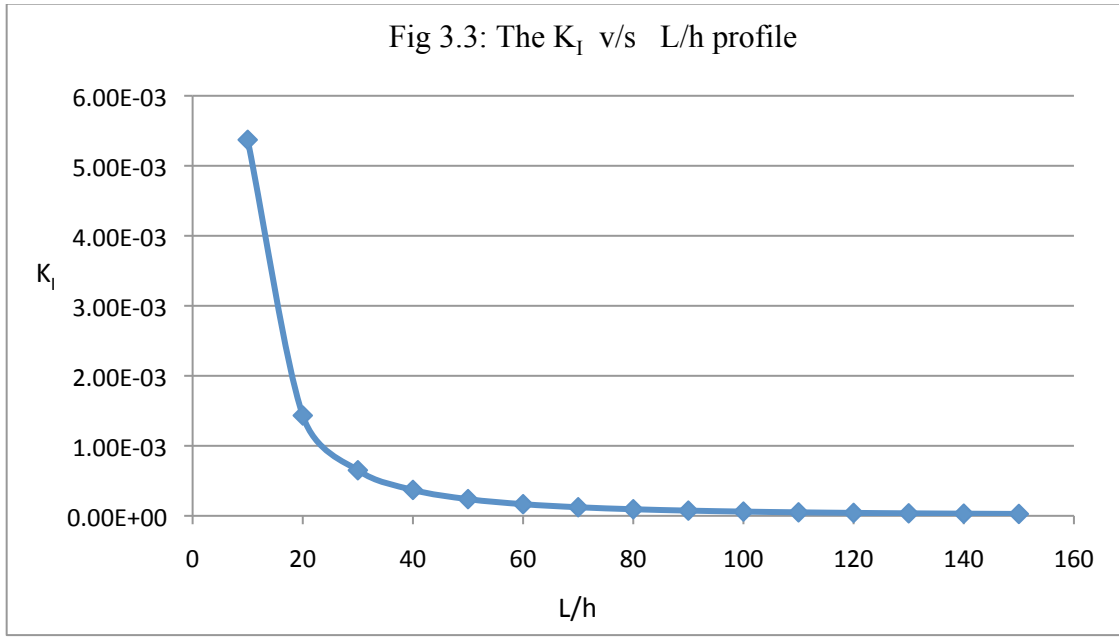

Fig 3.3: The $K_I$ v/s L/h profile

## 3.3 Analytical Solution for $K_I$:

In the present problem, we are considering the case of **constant displacement.** In this section, analytical solution of $K_I$ will be developed for this case.

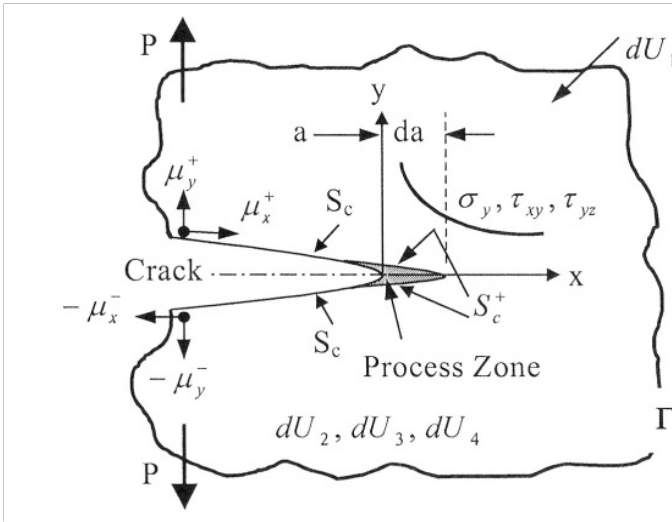

Fig 3.4: Cracked body with energy changes

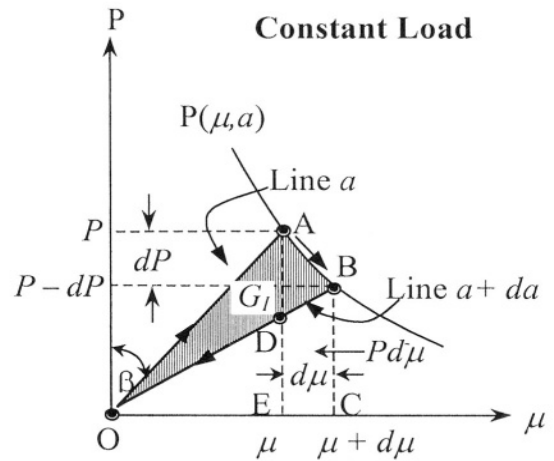

Fig 3.5: Load v/s Displacement diagram for a growing crack

Ref. to Fig 3.4:

$dU_1$ : The input energy change

$dU_2$ : Change in dissipated energy as heat.

$dU_3$ : Total potential elastic energy

$dU_4$ : Change in kinetic energy of the system

Consequently, the conservation of energy change due to the displacements arising from the fracture area change dA can be defined as:

$$\frac{dU_1}{dA} - \frac{dU_2}{dA} = \frac{dU_3}{dA} + \frac{dU_4}{dA}$$

Energy release rate $G_i, i = I, II, III$ is given as:

$$G_i = \frac{dU_1}{dA} - \frac{dU_3}{dA} = \frac{dU_1}{Bda} - \frac{dU_3}{Bda}$$

If the cracked plate shown in Figure 3.4 is subjected to an external load $P$ and the crack growth very slow, then the load-points undergo a relative displacement $d\mu$ perpendicular to the crack plane and the crack length extends an amount $da$. Consequently, the work done responsible for such an increment in displacement and crack length is defined by :

$$\frac{dU_1}{da} = P\frac{d\mu}{da}$$

Consider mode I (tension) loading and the linear behavior shown in Figure 3.5. The stored energy due to tension loading can be defined as the area under the curve.

$$U_3 = \frac{1}{2}P\mu$$

Hence, $\dfrac{dU_3}{da} = \dfrac{P}{2}\dfrac{d\mu}{da} + \dfrac{\mu}{2}\dfrac{dP}{da}$ and $G_I = \dfrac{1}{2B}\left(P\dfrac{d\mu}{da} - \mu\dfrac{dP}{da}\right)$.

Consider the present problem under **constant displacement.** In this case, the load and load gradient expressions are

$P = \dfrac{\mu}{C} = \dfrac{3\mu EI}{2a^3}$ ; $\dfrac{dP}{da} = -\dfrac{9\mu EI}{2a^4}$ ; $\dfrac{d\mu}{da} = 0$, Hence we get: $G_I = -\dfrac{\mu}{2B}\dfrac{dP}{da} = \dfrac{9\mu^2 EI}{4Ba^4}$ For $\mu =$ Constant

Since $I = \dfrac{1}{12}Bh^3$ ; $\dfrac{I}{B} = \dfrac{h^3}{12}$, we get: $G_I = \dfrac{3\mu^2 Eh^3}{16a^4}$

The SIF for Mode I loading is given by: $K_I = \sqrt{EG}$ ; $K_I = \sqrt{\dfrac{3\mu^2 E^2 h^3}{16a^4}}$ ; $K_I = \dfrac{\sqrt{3}\mu Eh^{\frac{3}{2}}}{4a^2}$

For the present problem, $\mu = 0.50$ ; $E = 1$ ; $h = 0.50$ ; a=75, Substituting these values, we get: $K_I = 1.3608\times 10^{-5}\ \dfrac{N}{mm^{3/2}}$, Numerical value of: $K_I = 2.70\times 10^{-5}\ \dfrac{N}{mm^{3/2}}$

Displacement and Stress Field:

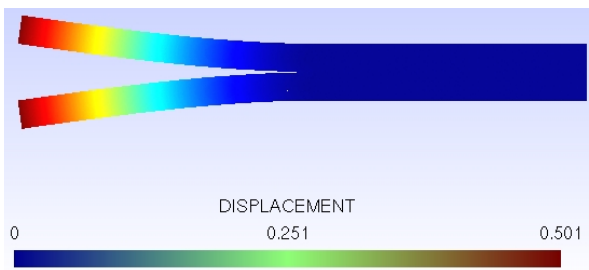
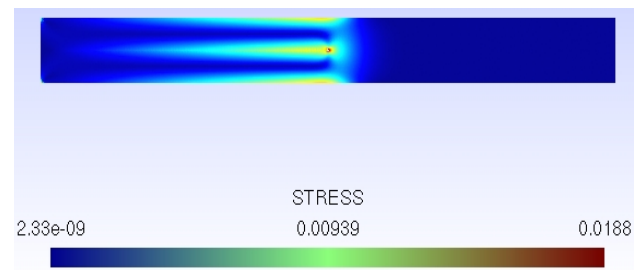

Comment:

- The analytical solution is developed for one dimensional beam problem. Hence the length of the beam is increased keeping the height constant until the L/h ratio is predominantly large i.e the geometry can be considered as one dimensional. However, the analytical solution is not exactly same as the numerical one as the numerical solution always associates different kinds of numerical error. However the order of magnitude is same.

- The enrichment zone used should be well inside the geometry. As shown in the adjacent figure, enrichment zone exceeding the geometry of the beam is not recommended.

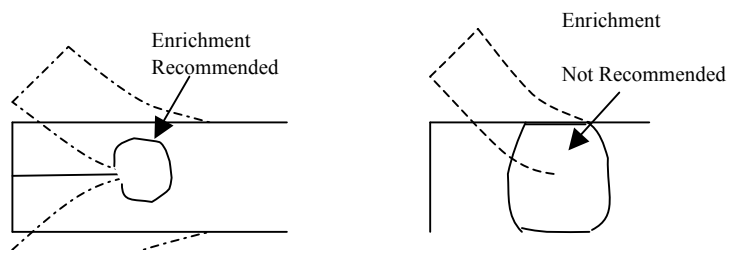

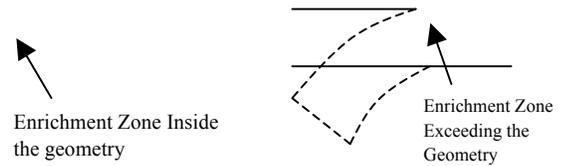
Enrichment Zone Inside the geometry    Enrichment Zone Exceeding the Geometry

# 4. Brazilian Test:

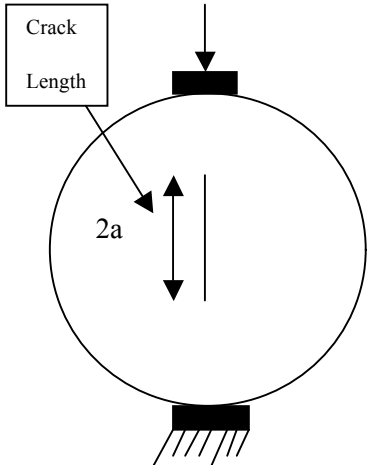
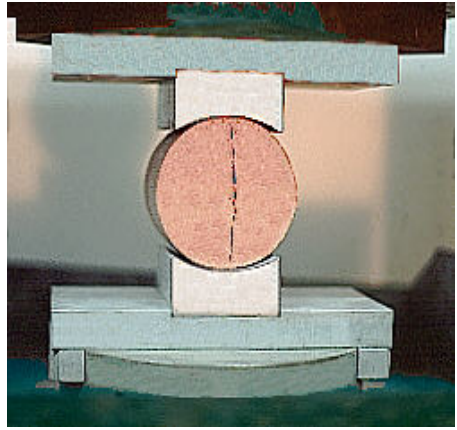
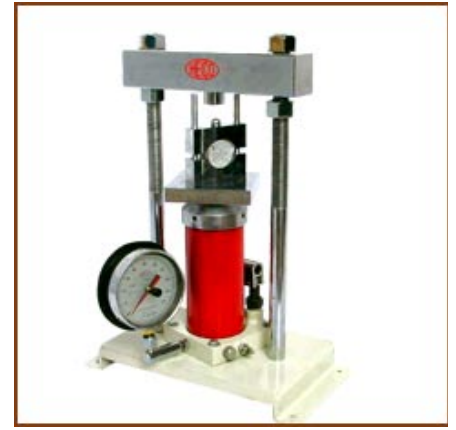

Fig 4.1: Schematic Diagram of the Setting of the Brazilian Test

Fig 4.2: The Setting of the Brazilian Test

Fig 4.3: The Experimental Set Up of the Brazilian Test

## 4.1 Problem Statement:

The Brazilian Test is a famous experiment on concrete sample, as represented on the figure. The goal of this exercise is to reproduce the experiment, where a crack appear in the center of the sample in the vertical direction and then propagate vertically until total failure of the sample. The displacement/force curve is to be plotted at the loading point while the crack propagates under the assumption that the crack propagate at constant value of $K_I$ stress intensity factor.

## 4.2 A Brief Background:

The Brazilian test was developed to measure the tensile strength of brittle materials like rocks and concrete (Berenbaum and Brodie, 1959). The Brazilian testing procedure is simple and the specimen preparation is easy compared to other test methods. Standard test method had been suggested (ISRM, 1978). The indirect tensile strength of a disc sample (Figure 1) of radius R and thickness t, with known load at failure P is given by

$$\sigma_t = \frac{P}{\pi R t}$$

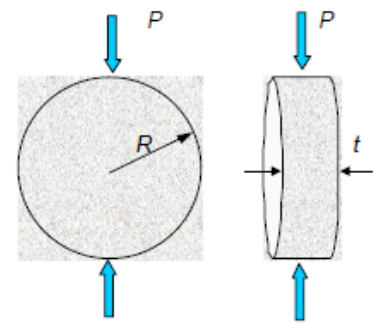

Fig 4.1: Brazilian test for indirect tensile strength

The stress field inside the disc can be obtained by solving a differential equation that employs Airy's stress function and satisfies the boundary condition of the sample.

Brazilian tests simulation of rock samples with pre-existing cracks is executed with the crack length and orientation taken as variables. Visible new cracks are generated right after the global peak in the load – displacement curve. It is seen that the macro tensile strength decreases as the pre-existing crack length increases.

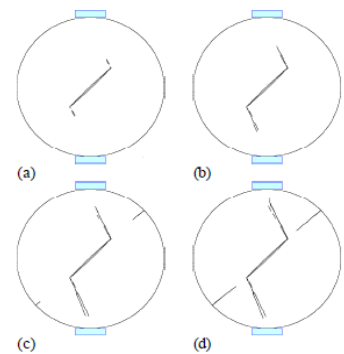

## 4.3: Test Sample and the Boundary Condition:

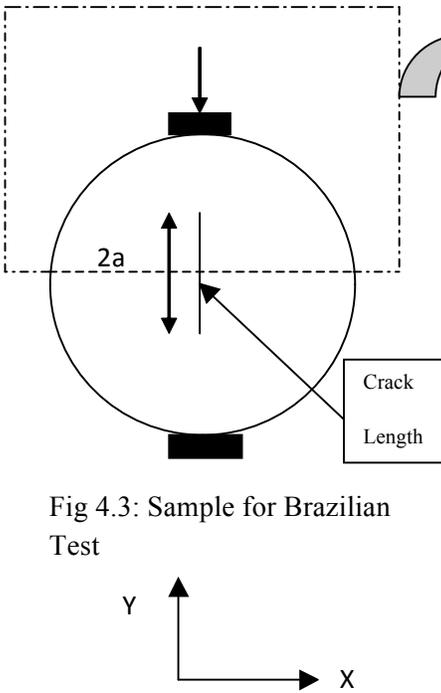

Fig 4.3: Sample for Brazilian Test

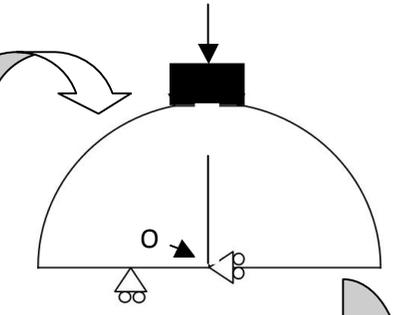

Fig 4.4: Test Sample before failure

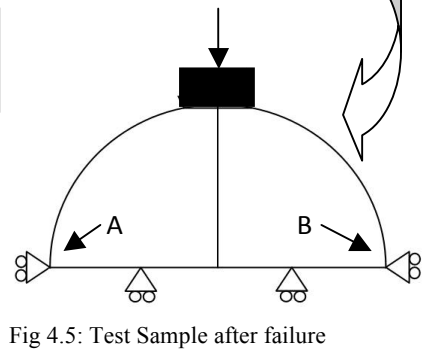

Fig 4.5: Test Sample after failure

**NOTE:**

The crack in the 3D sample is a penny shaped. The length of the crack is given by the diameter of the crack. Because of symmetry in loading and geometry, half of the sample is analyzed. The length of the crack is given by the radius.

<u>Boundary Condition:</u>

Until the crack propagation causing the failure of the sample, the BC is given by the Fig 4.4. The midpoint (O) motion is prevented in the X-direction and the motion of the other points is restricted in the Y-direction.

After the failure of the sample i.e. the crack propagates through the sample, the BC condition is given by Fig 4.5. Point A and B are restricted in the X-direction.

## 4.4: Methodology:

The SIF is related to the force as: $K_I = \alpha F$. The analysis is performed for a load of 1N. Hence, $K_I = \alpha$.

Now the critical force can be obtained as: $K_{Ic} = \alpha F_c$. The $K_{Ic}$ is taken as 1. Hence $F_c = \dfrac{1}{\alpha}$; Or, $F_c = \dfrac{1}{K_I}$.

Hence the obtained $K_I$ value is inverted to get the critical force. The obtained displacement from the simulation is for a force of 1N. Hence the true displacement is obtained by multiplying the obtained displacement with the $F_c$.

The sample with no crack will behave linearly (i.e. the Force v/s Displacement curve is linear). The sample will have higher modulus of elasticity. With the initiation of the crack, the sample will start losing the strength and will undergo snap back phenomenon. After the propagation of the crack through the body, the sample will still withstand load. But it will have the minimum strength.

The simulation if performed for the crack length from 0.10 to 0.95. The radius of the sample is 1. Hence a crack length > 1 stands for the full propagation of the crack and the failure of the material.

## 4.5: Results and Discussions:

Table 4.1 : Table for Force and Displacement

| Crack Radius | Displacement for F=1 N | $K_I$ | Critical Force | Actual Displacement |
|---|---|---|---|---|
| 0.1 | 5.31E-09 | 6.89E-02 | 1.45E+01 | 7.70E-08 |
| 0.15 | 5.33E-09 | 8.56E-02 | 1.17E+01 | 6.22E-08 |
| 0.2 | 5.35E-09 | 1.01E-01 | 9.89E+00 | 5.29E-08 |
| 0.25 | 5.38E-09 | 1.16E-01 | 8.61E+00 | 4.63E-08 |
| 0.3 | 5.42E-09 | 1.32E-01 | 7.60E+00 | 4.12E-08 |
| 0.35 | 5.46E-09 | 1.47E-01 | 6.79E+00 | 3.71E-08 |
| 0.4 | 5.52E-09 | 1.64E-01 | 6.11E+00 | 3.37E-08 |
| 0.45 | 5.60E-09 | 1.81E-01 | 5.53E+00 | 3.10E-08 |
| 0.5 | 5.69E-09 | 1.99E-01 | 5.03E+00 | 2.86E-08 |
| 0.55 | 5.80E-09 | 2.18E-01 | 4.60E+00 | 2.67E-08 |
| 0.6 | 5.94E-09 | 2.36E-01 | 4.23E+00 | 2.51E-08 |
| 0.65 | 6.11E-09 | 2.55E-01 | 3.92E+00 | 2.39E-08 |
| 0.7 | 6.31E-09 | 2.73E-01 | 3.67E+00 | 2.31E-08 |
| 0.75 | 6.55E-09 | 2.85E-01 | 3.50E+00 | 2.30E-08 |
| 0.8 | 6.83E-09 | 2.89E-01 | 3.46E+00 | 2.36E-08 |
| 0.85 | 7.16E-09 | 2.75E-01 | 3.63E+00 | 2.60E-08 |
| 0.9 | 7.51E-09 | 2.33E-01 | 4.30E+00 | 3.23E-08 |
| 0.95 | 7.87E-09 | 1.43E-01 | 6.97E+00 | 5.49E-08 |

## Force v/s Displacement Curve:

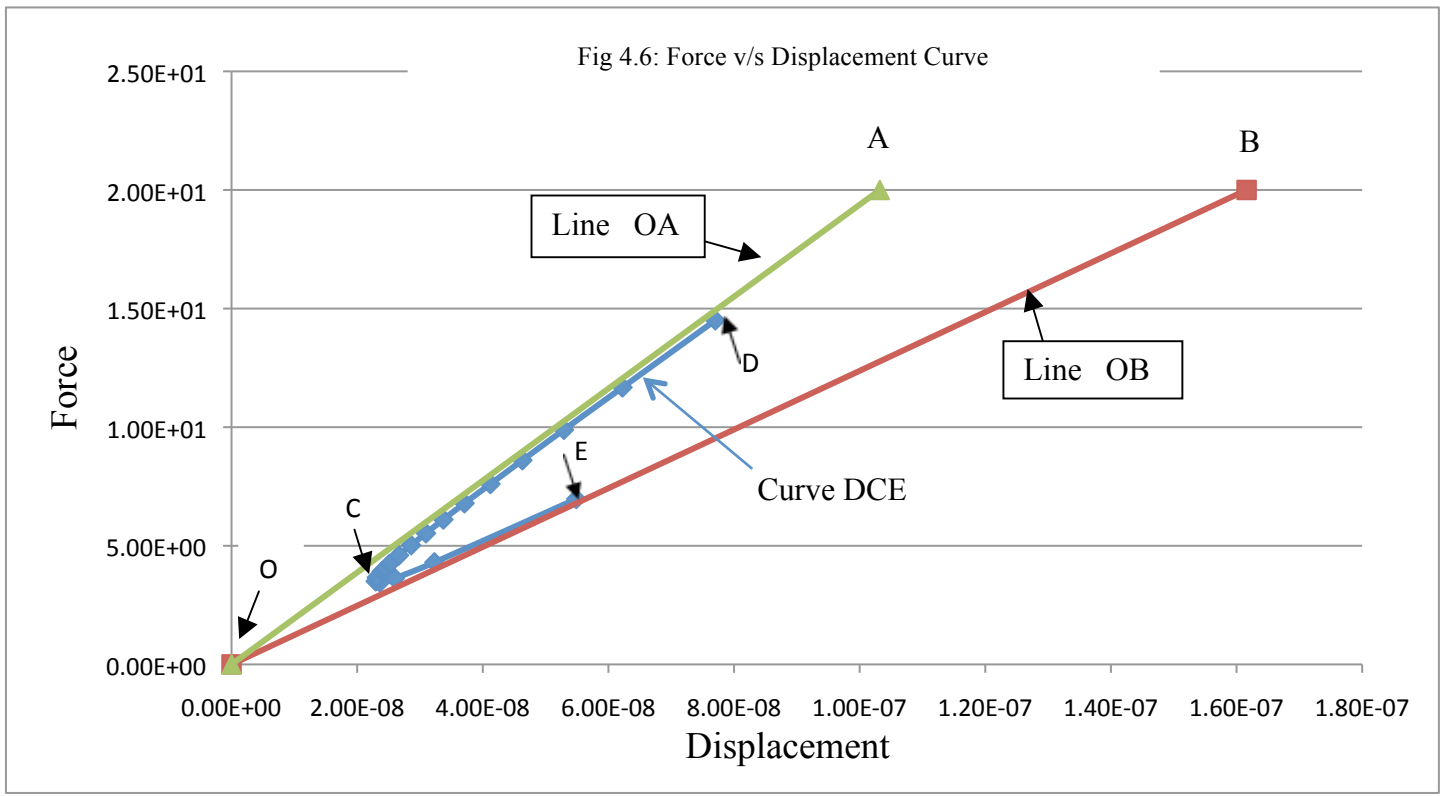

## Details of the Plotting of the Force- Displacement Curve (Ref. Fig 4.6):

### Curve DCE:

The numerical simulation is performed for different length of the crack. The obtained value of $K_I$ is inverted to get the actual force which when multiplied with the obtained displacement gives the true displacement. The obtained actual force and displacement is plotted to get the curve DCE.

### Line OA:

Table 4.2: Table for the Line OA.

| Displacement For F=1N in mm. | Displacement for Applied Force (mm) | True Applied Force (N) |
|---|---|---|
| 0 | 0 | 0.00E+00 |
| 5.16E-09 | 1.03E-07 | 2.00E+01 |

Line OA corresponds to the case of NO CRACK. Simulation performed for force of 1 N. The displacement (in the Y direction) is multiplied by 2 for an applied force of 2 N. The data in the last two columns of the table 4.2 are used to plot Line OA.

### Line OB:

Table 4.3: Table for the Line OB.

| Displacement For F=1N in mm. | Displacement for Applied Force (mm) | True Applied Force (N) |
|---|---|---|
| 0 | 0 | 0.00E+00 |
| 8.08E-09 | 1.62E-07 | 2.00E+01 |

Line OB corresponds to the case of COMPLETE CRACK PROPAGATION. Simulation performed for force of 1 N. The displacement (in the Y direction) is multiplied by 2 for an applied force of 2 N. The data in the last two columns of the table 4.3 are used to plot Line OB.

### Significance of the Force- Displacement Curve:

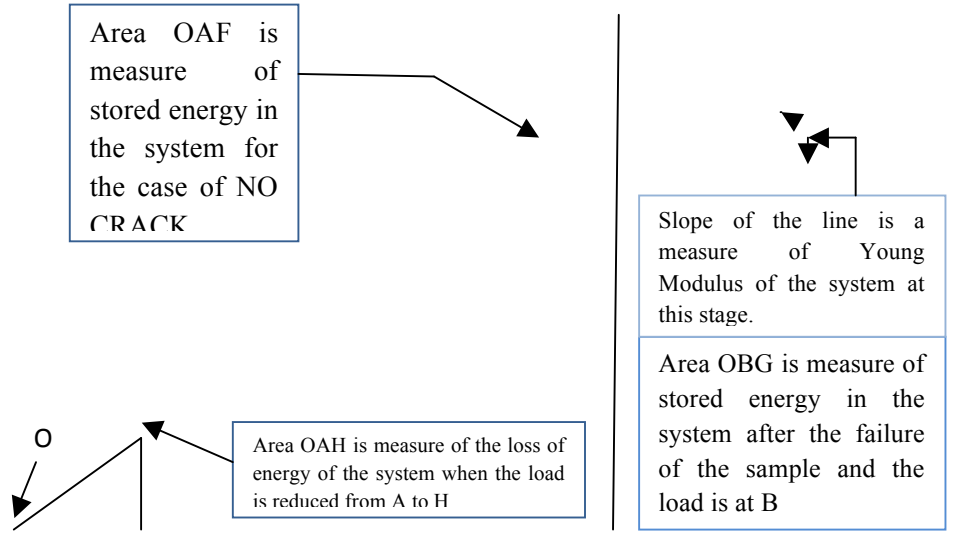

The force displacement curve is an example of **Snap Back phenomenon**. The are under the curve at a particular stage is a measure of the stored energy in the system. The difference of the area between the two stages gives a measure of the loss of energy of the system due to change in loading on the system. At a particular state, the slope of the line is a measure of the Young Modulus of the system.

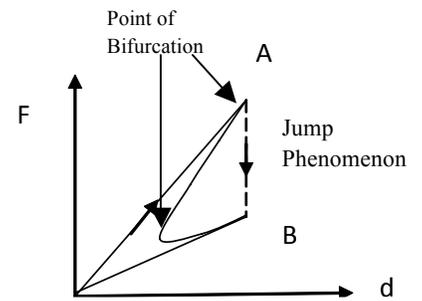

Fig 4.8: Jump Phenomenon

NOTE: In practice during experiment, it is difficult to capture the sudden change in the Force v/s displacement curve (the position of the global peak i.e. the **Point of Bifurcation**). The curve jumps suddenly from point A to point B . The phenomenon is known as **Jump Phenomenon.**

Displacement and Stress Plot:

| Crack Length | Displacement Plot | Stress Plot |
|---|---|---|
| 0 (NO CRACK) | | |

| Crack Length | Displacement Field | Stress field |
|---|---|---|
| 0.40 | 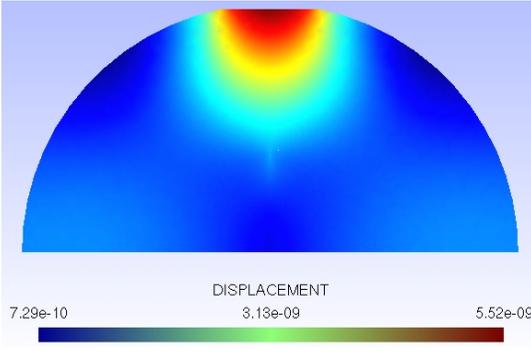 | 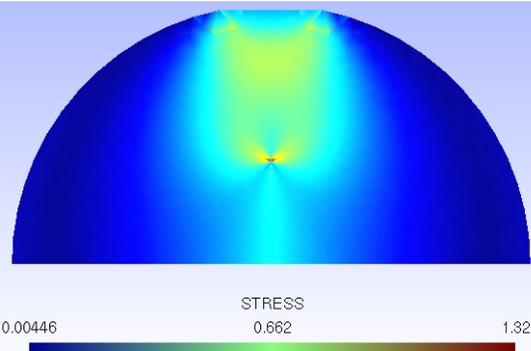 |
| 0.60 | 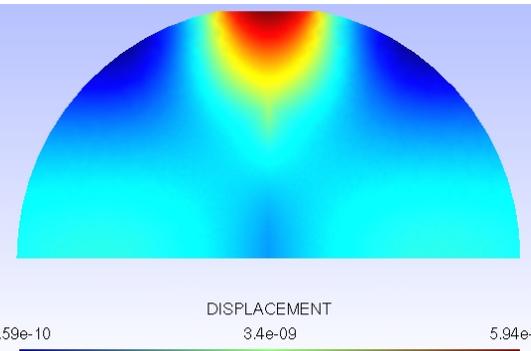 | 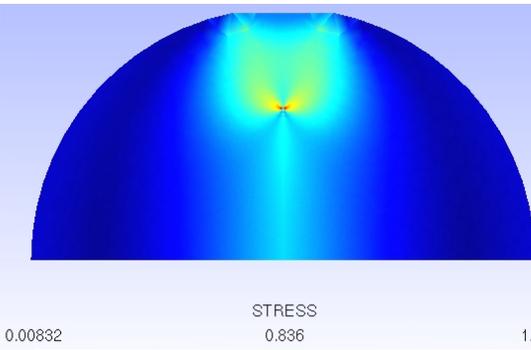 |

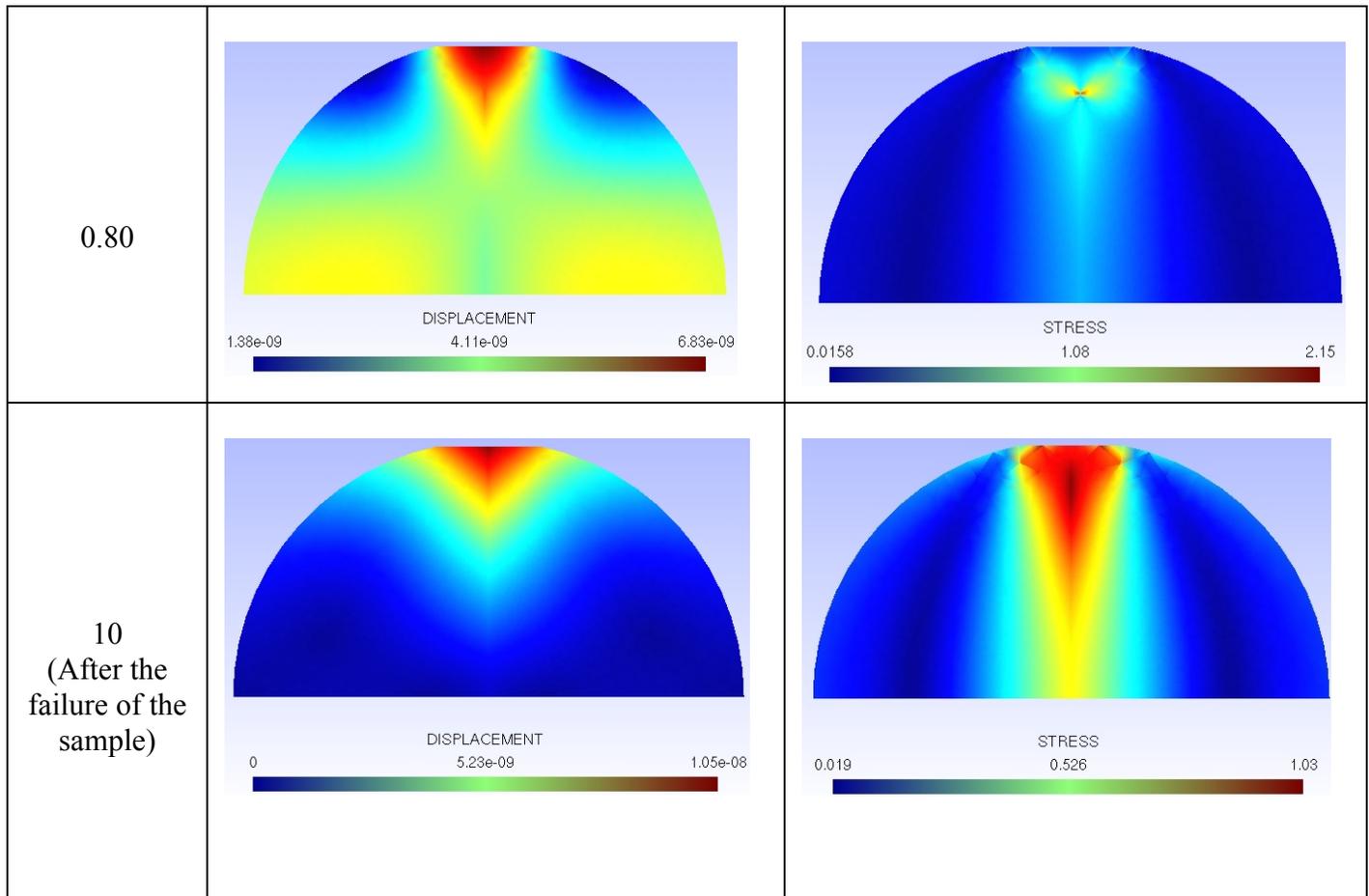

| | | |
|---|---|---|
| 0.80 | DISPLACEMENT 1.38e-09 4.11e-09 6.83e-09 | STRESS 0.0158 1.08 2.15 |
| 10 (After the failure of the sample) | DISPLACEMENT 0 5.23e-09 1.05e-08 | STRESS 0.019 0.526 1.03 |

## Comment:

- Brazilian Test gives an idea of the strength of the concrete specimen and its behavior under the uniaxial loading. The force at which the crack initiation occurs can be captured from the load displacement diagram.

- The displacement at the tip of the crack keeps on increasing with the increase of the crack length. Since the loading and the sample geometry is symmetric, the displacement field is also symmetric.

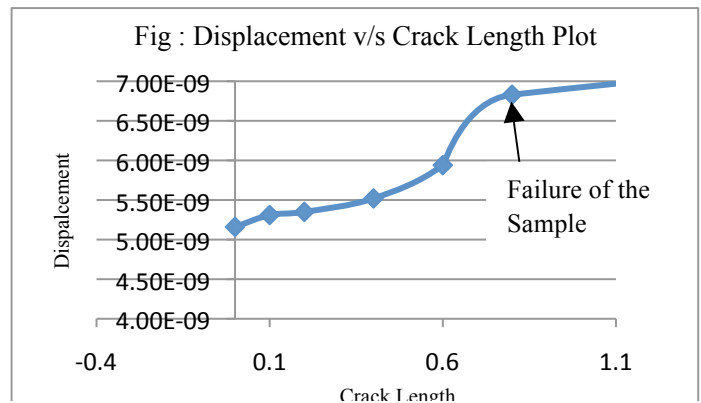

Fig : Displacement v/s Crack Length Plot

- Like the displacement field, the stress field is also symmetric due to the symmetry in loading and geometry.
  The stress is maximum at the tip of the crack. The stress is also reaches high value just beneath the loading plate.

Stress is also high just below the loading plate.

Stress is maximum at the crack tip (Theoretically Infinite)

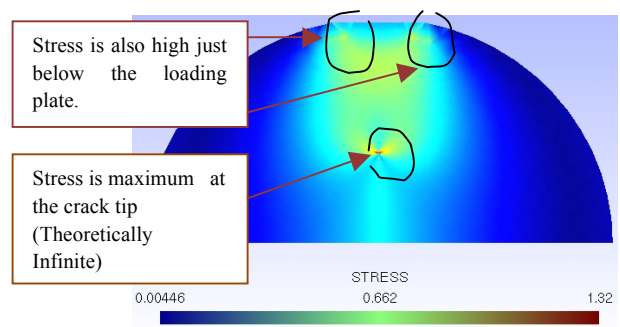

Fig: Stress Field at crack length 0.40

# 5. Conclusive Discussion:

- The study of convergence analysis gave an idea of the set of parameter that yields proper convergence. However the set of parameter should be optimally selected so as to optimize the speed of computation also.

- The convergence study can also be extended for the case of mixed mode, presence of hole, curved boundaries and other kinds of discontinuity which will simulate the real problems. The study can also be extended for other fields like stress, pressure, strain etc.

- The study of crack in a beam gives an idea about the SIF value when the 2-D model is enlarged enough to have 1-D behavior. Hence the study can be extended for the evaluation of SIF for 2-D model, for the mixed mode and for other discontinuities.

- The study of the Brazilian Test gives an idea of the Snap Back behavior of the sample under the specified loading condition. The Boundary Condition for the case of no crack and the failure of the sample should be properly chosen while analyzing the half of the sample.

- The study can be extended for the mixed mode and other types of loading condition (e.g. random loading, dynamic loading).

### Acknowledgement:

The author of this project report is highly indebted to **Prof. Nicolas Chevaugeon,** the instructor of the computer laboratory on X-FEM for his outstanding guidance throughout the session. It is only because of his soft guidance, the author was able to complete the project. The author is also highly indebted to **Prof. Nicolas MOES**, the instructor of the X-FEM module for his awesome instruction.